\documentclass[%
 aip,
 amsmath,amssymb,
preprint,%
]{revtex4-1}

\usepackage{graphicx}
\usepackage{xcolor}
\usepackage{dcolumn}
\usepackage{bm}

\usepackage[utf8]{inputenc}
\usepackage[T1]{fontenc}
\usepackage{mathptmx}
\usepackage{etoolbox}

\makeatletter
\def\@email#1#2{%
 \endgroup
 \patchcmd{\titleblock@produce}
  {\frontmatter@RRAPformat}
  {\frontmatter@RRAPformat{\produce@RRAP{*#1\href{mailto:#2}{#2}}}\frontmatter@RRAPformat}
  {}{}
}%
\makeatother
\begin{document}

\preprint{AIP/123-QED}

\title[Elastic modulus measurements of cooked Lutefisk]{Elastic modulus measurements of cooked Lutefisk}
\author{Blandine Feneuil}
 \affiliation{University of Oslo, Oslo, Norway.}
 \affiliation{Current adress: SINTEF Industry, Petroleum department, Trondheim, Norway.}
\author{Eirik Strøm Lillebø}
 \affiliation{Gamle Rådhus restaurant, Oslo, Norway}
\author{Christian Larris Honstad}
 \affiliation{Gamle Rådhus restaurant, Oslo, Norway}%
\author{Atle Jensen}
  \affiliation{University of Oslo, Oslo, Norway.}
\author{Andreas Carlson}
  \affiliation{University of Oslo, Oslo, Norway.}
  \email{\url{acarlson@math.uio.no} (A. Carlson), \url{blandine.feneuil@sintef.no} (B. Feneuil)}
  
\date{\today}

\begin{abstract}
\textit{Lutefisk} is a traditional Norwegian Christmas dish, made of dry cod soaked in a lye solution before re-hydrated. We report measurements of the tissue rheological properties of cooked \textit{Lutefisk}. Surprisingly, we find that the elastic modulus does not seem to depend heavily on cooking time, cooking temperature and the amount of salt, but depends instead mainly on the size of the fish fillets and the period of fishing. Although the salting and cooking of the \textit{Lutefisk} affects strongly the visual aspect of the fish fillets, these changes are not measurable with a rheometer. 
\end{abstract}

\maketitle

\section{Introduction}

\begin{figure*}[!ht]
\begin{center}
\includegraphics[height = 5cm]{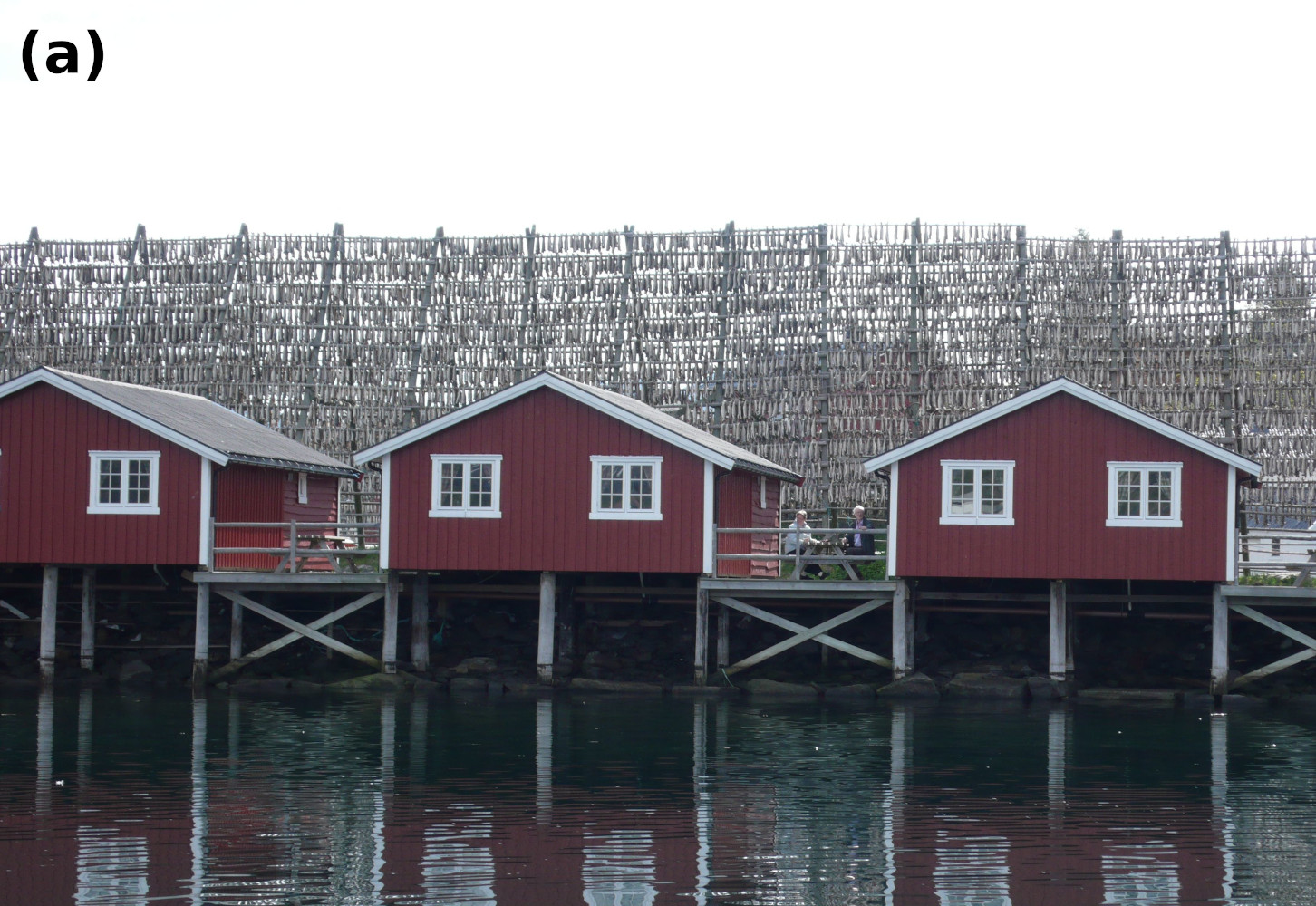}
\includegraphics[height = 5cm]{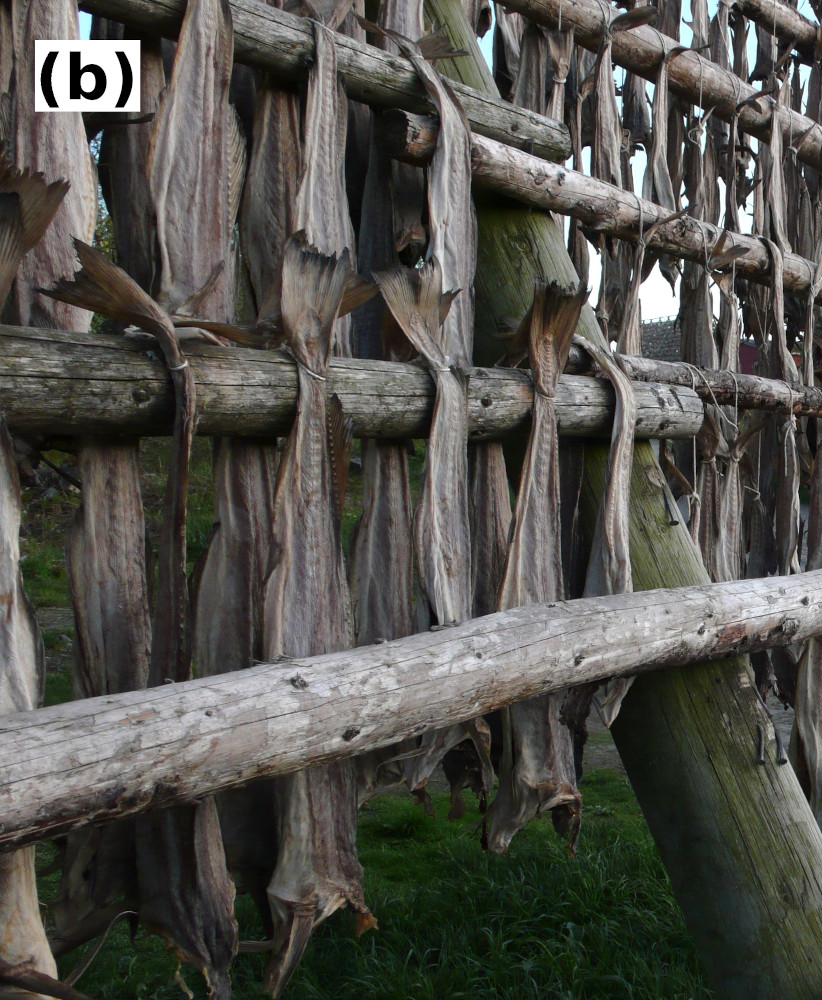}
\caption{(a) The wood racks, where the cod is hung in spring, are an important part of the landscape in Lofoten. (b) Dry cod on the racks. Both pictures have been taken by Atle Jensen in Lofoten, Norway, in May 2009.}
\label{figure_intro}
\end{center}
\end{figure*}

\textit{Lutefisk} is one of the most mythical dishes in Norway and an essential dish during Christmas holidays in Scandinavia. It is prepared from fish, usually cod, which is fished in the late winter and hung on wood racks for several months (see Fig. \ref{figure_intro}). Exposure to sun and wind makes it loose about 70\% of its initial weight, and this dried fish is known as stockfish \cite{2000_Johansen}. ``Lut" is the Norwegian word for ``lye", referring to the rehydration method of the stockfish used to prepare \textit{Lutefisk}: it is soaked in water for several days, then in a lye solution a few days, and finally rinsed in water. This process alters the tissue properties of the fish, where we in this work focus on measuring the elastic modulus of the fish flesh and how it changes during cooking.

The \textit{Lutefisk} tradition is accompanied with several myths about its discovery. One version is that it originated from a viking village where the stockfish resources were burnt down by a rival clan. Heavy rain extinguished the fire, and the wet ashes produced a lye-like solution, where the stockfish laid for several days until it was discovered by the hungry villagers. The fish did probably not look appetizing, but as they did not have anything else left to eat, they washed the fish, cooked it, ate it... and enjoyed it ! Another history links the discovery to a poor family that had let their fish fall in a washing pot \cite{2011_Johansen}. In reality, the origin of \textit{Lutefisk} is rather unclear. Who was the first to soak stockfish in a lye solution, and whether this event was accidental, is unknown. First sources on \textit{Lutefisk} date from 1540 in Sweden and 1770 in Norway \cite{2011_Johansen, 1990_Riddervold}. On the one hand, drying has been used for several thousands years to preserve fish \cite{2011_Johansen} and other meats \cite{2009_Belitz}. The fish enzymes that metabolise food still continue working after death, leading to the degradation of the fish and to the characteristic rotten fish smell \cite{2000_Barham}. The enzyme activity is stopped in the absence of water, so that dry fish can be kept all along the winter months and used as food in long expeditions. On the other hand, lye curing of food is known since at least the XVI century, and is still today used in the food industry, for instance to remove the bitterness of fresh olives \cite{2005_Chammem,2008_Maldonado,2009_Belitz}.

\textit{Lutefisk} is for hobby cooks known to be notoriously hard to cook to perfection \cite{2011_Johansen, 2019_Lin}. 
Most times, we believe, the end result becomes an overcooked gel-like consistency lacking any resemblance to fish meat. To better understand what affects the properties of the fish meat during cooking, we deploy a rheology methodology to measure the material properties of cooked \textit{Lutefisk}. Texture of food is directly related to how the consumer appreciates it: food must be easy to chew and provide a pleasant feeling in the mouth. In particular, the texture of cod has been the object of several studies \cite{1974_Love,1979_Dunajski,2003_Barat,2011_Skipnes,2019_Blikra}. The authors have shown that the softness/hardness is affected by the cooking temperature \cite{1979_Dunajski,2011_Skipnes,2019_Blikra}, the salt content \cite{2003_Barat} as well as  the water content, the pH and the fish size \cite{1974_Love}. In these studies, several different methods were chosen to evaluate the texture, such as chewing by a panel of persons \cite{1974_Love}, compression experiment \cite{2011_Skipnes} and shear experiment \cite{2019_Blikra}. 

To the best of our knowledge, no texture study has been published in a scientific journal on \textit{Lutefisk}. Nowadays, the numbers of \textit{Lutefisk} consumers across the world is quite limited, the tradition being preserved mainly in Norway, Denmark and Sweden, as well as Scandinavian families in the US. This can explain that, despite the intriguing and fascinating nature of the dish, it has been the object of very few scientific studies. We have found in the literature two papers, on the microorganism content of \textit{Lutefisk} \cite{2018_Lunestad}, and on spectroscopy experiments perform used to compare different commercial \textit{Lutefisk} brands \cite{2020_Hassoun}. The peculiar aspect of the dish has raised our attention, and we have focused on quantifying the consumer experience with scientific tools. We use a rheometer to measure the elastic shear modulus, a material property that can be easily compared with others publications and other materials. 

Here, we have to mention that relating the rheological properties of food with consumer perception is a difficult task \cite{1977_Moskowitz,2019_Yoshida}. The properties measured with a rheometer are in general not enough to quantify the wide range of personal impressions that food can provide, such as soft, firm, tender, crumbly, pasty, juicy and sticky. Indeed, while a rheometer provides a measure that is averaged on the whole sample, the consumer tongue is sensitive to the heterogeneity of the food \cite{2021_Thomazo}. Fish flesh is composed of muscle fibers surrounded by a connective tissue \cite{1979_Dunajski, 2000_Barham}. Contrarily to mammals whose muscle cell can be 15 cm long, fish fibers are short and separated by membranes called myocommata (see Fig. \ref{figure_samples} (b,c)). The rheological properties of fish is therefore expected to result form the properties of the muscle fibers, the connective tissue and the myocommata.

However, in some specific cases, rheological measurements have been successfully related to food preparation. The storage shear modulus of cod flesh \cite{2019_Blikra} and surimi \cite{2017_Murthy} was found to increase during cooking when the internal temperature of the sample was between 50 and 70$^o$C. The rheological properties of gelled desserts has been compared with the consumer appreciation \cite{2019_Yoshida}, and it was observed that the elastic properties and the yield stress should neither be too large or too small.

In the following, we present measurements of the elastic shear modulus of \textit{Lutefisk}. The \textit{Lutefisk} fillets have been prepared following the recipe used in the \textit{Gamle Rådhus } restaurant in Oslo. The placement in the fish, the cooking time, temperature and the salt amount have been varied in order to measure their effect on the shear modulii. These experiments are compared to the chef's evaluation of the cooking level in order to check whether the rheological measurement is an adequate tool to assess the proper cooking time of \textit{Lutefisk}.

\section{Materials and methods}

\subsection{Sample preparation}

To obtain consistent result, we use \textit{Lutefisk} pieces from one producer: Brødrene Berg from Værøy, Norway. The fish is received in the form of fillets, which have already been cured in lye. The fillets selected by Brødrene Berg for use at the \textit{Gamle Rådhus} restaurant in Oslo, that specializes on this dish, are made only from the highest quality of stockfish, i.e. the largest fishes without defects \cite{2011_Johansen}. Two major types of pieces can be distinguished, as illustrated in Fig. \ref{figure_samples}. The larger pieces, which were located close to the fish head, have thicknesses comprised between 40 and 55~mm, length between 120 to 150 mm, and initial mass between 420 and 570~g. Fish-tail pieces are 20 to 40 mm thick, 60 to 130 mm long and weight between 250 and 470~g (see Fig \ref{figure_samples}(a)).

All the fillets are first washed, then salted. After salting, pieces were left at rest during 45 to 100 minutes before cooking them in a steam oven. We have checked that increasing the salting time from 45 to 100 min does not affect the mass or the elastic modulus. When the fish is served at the restaurant \textit{Gamle Rådhus}, it is cooked 12 minutes at 100$^o$C in a steam oven. In our experiments, cooking temperature and duration have been varied. The salting and cooking lead to important losses of water from the samples. Weight loss is evaluated by weighting the samples before salting, just before cooking and after cooking. The pH of the samples has been measured using pH-paper, and a constant value around pH$\simeq$10 has been obtained for all samples, before and after cooking. The fillets are left to cool at ambient temperature, then slices are cut, set into petri dishes and stored at 100\% humidity and at room temperature (about 20$^o$C) until testing within 12h.

We have performed three series of experiments, following the same procedure. The first two series took place in the advent time in 2019. Advent time is the main \textit{Lutefisk} season. The experiment series 3 was performed in October 2021, using the first delivery of \textit{Lutefisk} of the season that could be provided by the producer. The first two series give comparable results and have been used to study the effect of temperature and cooking time. The third series has focused on the the effect of salt content.

\subsection{Rheometry}

\begin{figure*}[!ht]
\includegraphics[width = \textwidth]{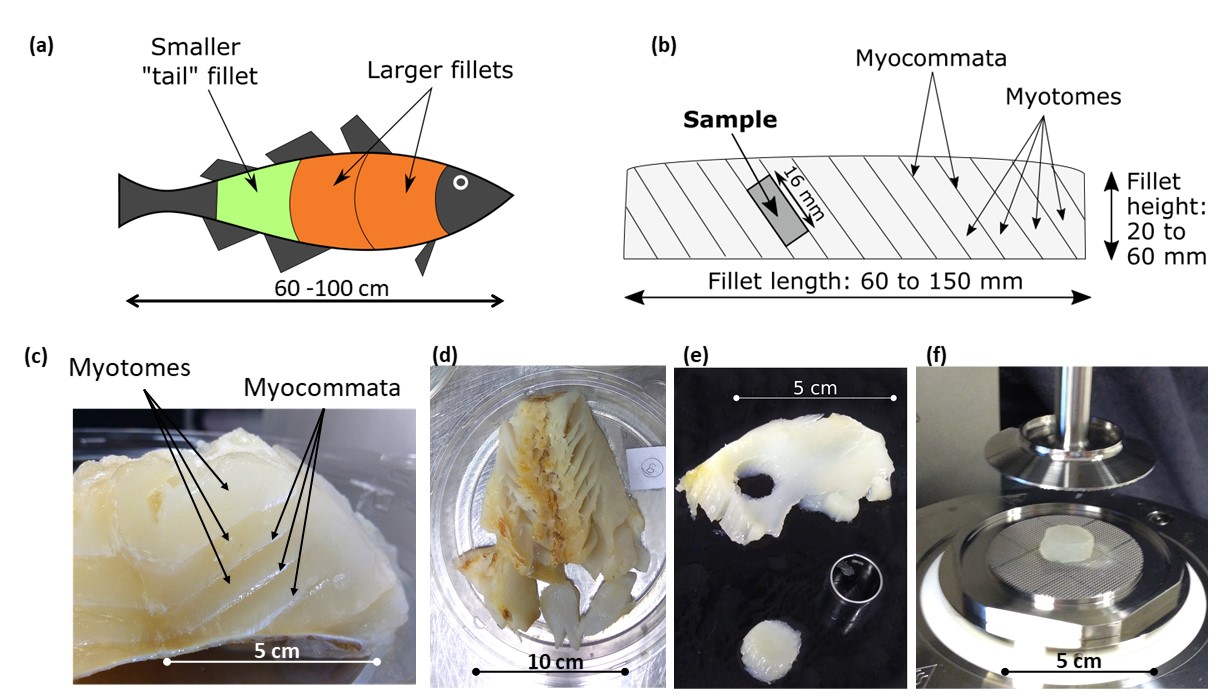}
\caption{ (a) Sketch of the cod with the location of the two types of \textit{Lutefisk} fillets used in this study: smaller "tail" fillets and larger fillets. (b) Structure of a cod fillet, where layers of muscle fibers (myotomes) are separated by a membrane (myocommata), and schematic representation of the location of a rheometry sample in the fillet. (c) Cross section of a salted uncooked sample, where the myocommata membrane can be seen. (d) Top view of a cooked fillet, where the myocommata has broken and flakes appear. (e) Each 16 mm - diameter sample is cut from a myotome using a cookie cutter. (f) A sample placed in the rheometer at the center of the rough plate-plate geometry.}
\label{figure_samples}
\end{figure*}

The flesh of the fish is composed of myotomes (muscle fibers) separated by a white membrane (the myocommata). The myocommata is harder to cut than the myotomes in the raw fish. During cooking the membrane breaks down and the myotomes form flakes (see Fig. \ref{figure_samples}(c) and \ref{figure_samples}(d)). In order to compare the different fillets and preparation methods, we have taken each rheometry sample in one single myotome, avoiding to include the myocommata. The samples are cut with a circular cookie cutter, as shown in Fig. \ref{figure_samples}(e). The sample diameter must be smaller than the myotomes width, but as large as possible to be representative of the average fish flesh. We choose 16 mm diameter as a compromise. The samples are taken at the middle height of the myotomes in order to study the properties of the middle of the fish fillets (see Fig. \ref{figure_samples}(b)).

The disk-shaped sample is then placed at the center of a 50-mm diameter plate-plate rheometer geometry (see Fig. \ref{figure_samples}(f)). Wall slip of soft solid on the rheometer plates must be prevented during rheometry measurements \cite{2021_Kamkar}. To address this issue, we use striated plates, with roughness size 0.5 mm. This size is larger that the diameter of fish fibers ($\sim$ 100 $\mu$m \cite{2006_Johnston}). Temperature of the plates is regulated with a thermostat at 20$^o$C. The rheometry test is carried out as  fast as possible after the sample is removed from the 100\% humidity box in order to avoid drying.

A constant normal force of 2N is applied on the sample during the test: this force is enough to hold the sample without breaking it. Storage (i.e. elastic) and loss (i.e viscous) modulii of the sample are obtained by oscillatory rotation of the upper plate. For all samples, we applied at constant angular frequency 10 rad/s and strain amplitude increasing from 0.0032 \% to 3.2\%. For some samples, we also perform complementary tests at strain amplitude 0.16\% and angular frequency decreasing from 100 to 0.1 rad/s. 

As the plate-plate geometry is not properly filled with the fish, the strain and stress values are not those provided by the rheometer software, but they must be calculated from  the rotation angle and the torque. Conversion relations between the torque T and the stress $\tau$, and between the rotation angle $\varphi$ and the strain $\gamma$ are 

\begin{eqnarray}
\tau & = & \dfrac{3}{2} \dfrac{T}{\pi R^3},\\
\gamma & = & \dfrac{3}{4} \dfrac{\varphi R}{H},
\end{eqnarray}
where R = 8 mm is the sample radius and H is the distance between the plates, which is measured for each sample. In order to use these equations, we need to check that the inertia of the rotating plate is small enough, i.e. the contribution to the measured stress due to inertia is negligible compared to the stress related to the sample \cite{2015_Ewoldt, 2017_Hudson}. This point will be discussed in paragraph \ref{part_shape_curves}. For each \textit{Lutefisk} fillet, four samples were cut and tested, where we present the average value of the elasticity, and the maximal and minimal values are indicated by the error bars (see results below).

\section{Results and discussion}

\subsection{Mass loss}

\begin{figure}[!ht]
\begin{center}
\includegraphics[width = 0.46\textwidth]{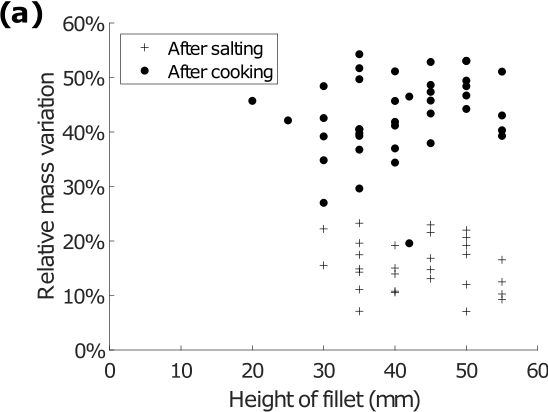}
\includegraphics[width = 0.46\textwidth]{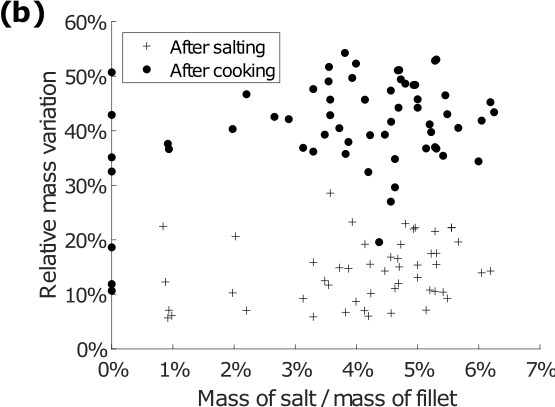}
\caption{Mass losses after salting and resting for at least 45 min, and total mass loss after salting and cooking. (a) Plot for series 1 and 2 as a function of the fillet height. (b) Plot for all three series as a function of the relative salt mass. Weight loss is evaluated by weighting the samples before salting, just before cooking and after cooking. Series 1 and 2 refer to samples prepared in the main \textit{Lutefisk} season, while series 3 corresponds to cod fished in the early season.}
\label{figure_mass_loss}
\end{center}
\end{figure}

During the preparation and cooking of the \textit{Lutefisk} fillets, water is released from the fish, leading to major weight losses. The first weight loss occurs after the fillets are salted, when they are left at rest for 45 min or more. Then, additional water is released from the fillets during cooking. We plot in Fig. \ref{figure_mass_loss} the mass loss caused by salting, and the total mass loss after salting and cooking. The mass variation due to the addition of salt before cooking was between 8 - 25\% of the initial weight. Depending of the samples, the time between salting and cooking varied from 45 min to 100 min. We have checked that the average value of the mass loss does not depend on the resting time, showing that the amount of water released after 45 min is small. During cooking, mass loss increases up to 35 to 55\%. Unsalted samples reach the same final mass loss as the salted samples if the cooking time is long enough ($\gtrsim$ 10 min), whereas for cooking time shorter that 10 min, mass loss of 10\% have been measured. We have attempted to understand the discrepancy between the mass losses of the samples by plotting them as the function of the fillet thickness and the salt amount in Fig. \ref{figure_mass_loss}. None of this parameters could be correlated with the mass loss.

\subsection{Shape of rheological curves}
\label{part_shape_curves}

\begin{figure}[!ht]
\begin{center}
\includegraphics[width = 0.46\textwidth]{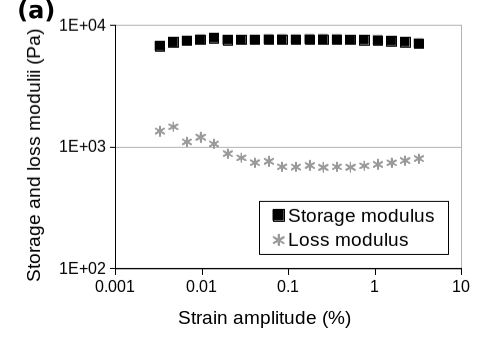}
\includegraphics[width = 0.46\textwidth]{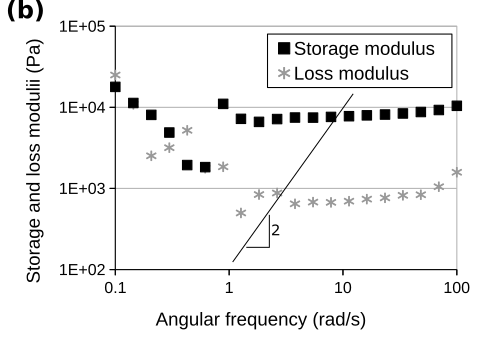}
\caption{Example of storage and loss modulii curves measured by (a) amplitude sweep at 10 rad/s and (b) frequency sweep at strain 0.16\%. The line indicates the slope of the G' curve in the case where the inertia of the plate is dominant compared to the sample properties.}
\label{figure_curve_shape}
\end{center}
\end{figure}

Two types of oscillation tests have been carried out, as illustrated in Fig. \ref{figure_curve_shape}. When the angular frequency of the oscillations is $\omega = 10$ rad/s, the storage modulus $G'$ is nearly constant when the strain amplitude varies from $3.2 \cdot 10^{-3}$ \% to 3.2 \%. This confirms that no slip appears at the larger oscillation amplitudes. The loss modulus $G''$ decreases slightly in this range and remains one order of magnitude smaller than the storage modulus. 

During a frequency sweep at a constant strain amplitude ($\gamma = 0.16\%$), both $G'$ and $G''$ increase from 1 rad/s to 100 rad/s. At lower angular frequency, the rheometer measurement becomes inaccurate. The contribution of the tool inertia to the apparent elastic modulus is proportional to the square of the frequency\cite{2015_Ewoldt,2017_Hudson}: $G'_{Inertia} \propto \omega^2$ . The increase of the measured elastic modulus from $\omega = 1$ to 100 rad/s, is much lower than this contribution (Fig. \ref{figure_curve_shape}(b)). Therefore, the instrument inertia is negligible if the angular frequency is low enough, in particular at the conditions that are used in the following of this paper: angular frequency $\omega = 10$ rad/s and amplitude $\gamma = 0.16\%$.

For all the tests at fixed $\omega$ and $\gamma$, the phase angle d, defined by $\tan d = G''/G'$, is constant for our experiments: $\tan d = 0.1 \pm 0.02$ for series 1 and 2 and $\tan d = 0.08 \pm 0.02$ for series 3; these values are not affected by salting and cooking. This means that the variations of $G'$ and $G''$ are similar, and it is sufficient to discuss the values of the storage modulus $G'$.

\subsection{Elastic modulus, cooking time and fillet size}

\begin{figure}
\begin{center}
\includegraphics[width = 0.46\textwidth]{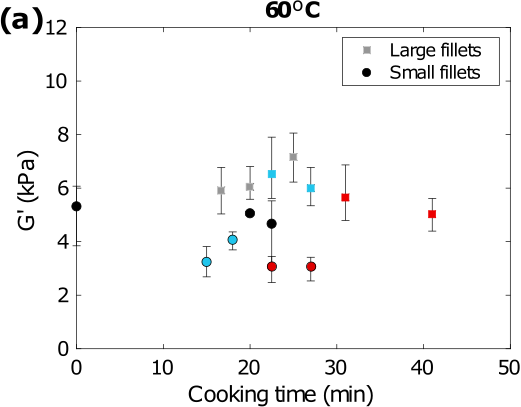}
\includegraphics[width = 0.46\textwidth]{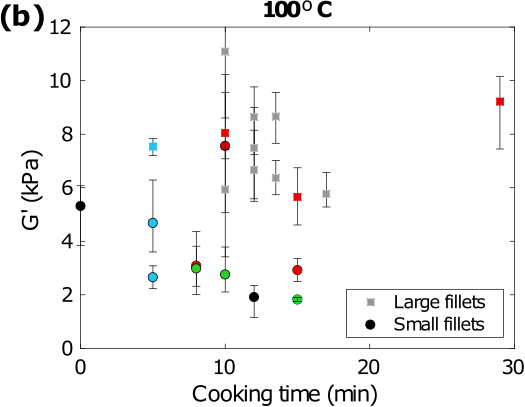}
\includegraphics[width = 0.46\textwidth]{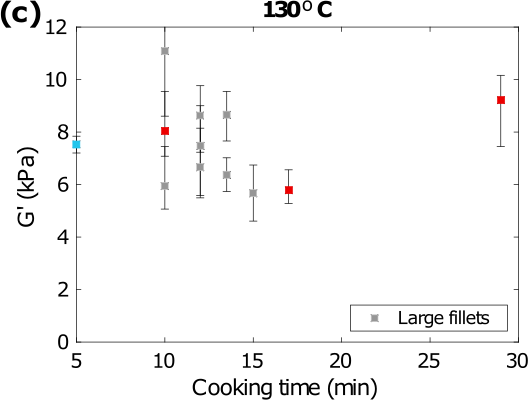}
\caption{Storage modulus at 10 rad/s and 0.16\% amplitude, for experiment series 1 and 2 and three cooking temperatures: (a) 60$^o$C, (b) 100$^o$C and (c) 130$^o$C. Black symbols indicate large samples and grey symbols, "tail" samples. Error bars indicate the smaller and larger values measured on four different pieces cut from the same fish fillet. The colors indicate the visual cooking assessment of the fillets: blue: undercooked (blue), well-cooked (red) and overcooked (green). Grey and black points refers to fillets where the cooking aspect has not been assessed.}
\label{figure_results}
\end{center}
\end{figure}

Storage modulus experiments in series 1 and 2 are shown in Fig \ref{figure_results}. For all these experiments, the large fillets had received 26~g of salt and the small fillets, 10~g of salt, in conformity with the restaurant's recipe. First, we note that although experiment series 1 and 2 have been performed on different days, separated by a few weeks, they give similar values and trends. This verifies our experimental protocol.

Then, we observe that the cooking time does not have a major effect on the results: for each temperature and sample size, no increasing or decreasing trend can be noted when the cooking time increases. This comes to a rather great surprise to us as we can clearly distinguish visually an evolution of the aspect of the samples. This point will be further discussed later. Cooking temperature does not seem to affect $G'$ either. Note that we did not observe any correlation between the water loss and the storage modulus (results not shown here).

On the contrary, the type of the piece has a major influence on the fish elasticity: storage modulus of the small "tail" pieces is always smaller than the modulus of the large fillets. Most values are between 6000 Pa and 10000 Pa for the large samples, and between 2000 and 6000 Pa for the "tail" samples.

Our results can be compared with measurement on non-cured cod muscle. \textit{Blikra et al.} \cite{2019_Blikra} measured the elastic modulus of cod samples during heating from 0 to 100$^o$C. The modulus of the raw cod was 22 000 Pa, this is several times larger than our measurements on raw \textit{Lutefisk}. This shows that the drying, water and lye treatment of the \textit{Lutefisk} makes its softer than regular fish.  Besides, \textit{Blikra et al.} report first a small decrease of the storage modulus when the temperature rises to 40$^o$C, then an increase to 48 000 Pa at 80$^o$C. Such variations are not observable when cooking \textit{Lutefisk}. We believe that lye denatures the collagen or other proteins present in the fish muscle, i.e., their structured shape is broken. In order to fully characterise the chemical effect of the lye on the proteins, an additional chemical analysis would be needed, which is beyond the scope of this paper.

\subsection{Effect of salt content}

\begin{figure}[!ht]
\begin{center}
\includegraphics[width = 0.46\textwidth]{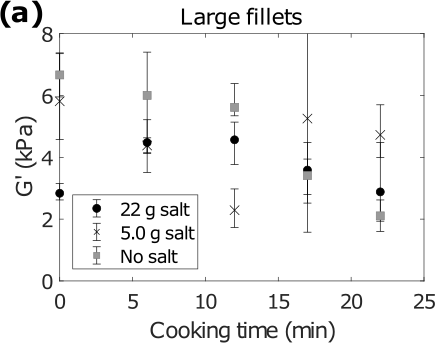}
\includegraphics[width = 0.46\textwidth]{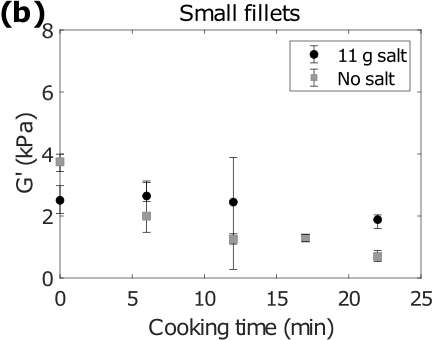}
\caption{Elastic modulus of samples in series 3, where the salt amount has been studied, for (a) the large fillets and (b) the smaller "tail" fillets. Error bars indicate the smaller and larger values measured on four different samples cut from the same fish fillet.}
\label{figure_salt}
\end{center}
\end{figure}

For the third series of experiments, we have decreased the amount of salt from the standard value to zero. Temperature is kept constant at 100$^o$C. Results are given in Fig. \ref{figure_salt} for both types of fillet. First, we observe by comparing these results with Fig. \ref{figure_results} that the measured $G'$ values are consistently smaller in series 3 than in series 1 and 2. In series 3, 2000~Pa $\leq G' \leq$ 7000~Pa for the larger fillets, and 700~Pa $\leq G' \leq$ 4000~Pa for the smaller "tail" fillets. Therefore, the cod which is fished and prepared in the early season appears to give a softer \textit{Lutefisk} than in the main season. However, the same trends can be observed for all series of experiments: no major change of elasticity occurs during cooking, and $G'$ is smaller for smaller "tail" fillets than for larger fillets. 

The new result from Fig. \ref{figure_salt} is that the salt content does not affect the elastic modulus. This result is surprising because the visual aspect of the samples is clearly dependent on the presence of salt. Unsalted samples have a gelly-like transluscent aspect, whereas salted fillets are white. Therefore, $G'$ measurement does not reflect the strong visual differences between the samples.

\subsection{Cooking assessment}

In order to discuss the relation between $G'$ and the aspect of the \textit{Lutefisk}, i.e. the appreciation of the dish by a consumer or a cook, we have assessed the cooking level of the fillets on a 3-step scale: "undercooked", "well-cooked" and "overcooked". The assessment is carried out on the whole fillets, i.e before cutting the small disk-shaped samples, for experiment series 2. The fillets are observed by eye, and sometimes additionally touched with a finger or a tool. The evaluations are indicated by the color points in the graphs of Fig.  \ref{figure_results}.

First, we observe that the large fillets never seem to overcook, neither by increasing the cooking temperature or by doubling the cooking time explored here. Small samples, on the other hand, were overcooked when they remained in the steam oven at 100$^o$C for more than about 10 min. Our experiments allow us to suggest some tips for the preparation of \textit{Lutefisk}. We can first note that, although recommended cooking temperature in a steam oven is 100$^o$C, well-cooked pieces can be obtained for all cooking temperatures, provided that the cooking time is adjusted. More importantly, we have noted that the best final aspect of the small fillets was obtained at 60$^o$C. This suggest that for optimizing cooking of small pieces of \textit{Lutefisk}, lowering the cooking temperature is recommended.

\bigbreak

The results in Fig. \ref{figure_results} show that the cooking quality cannot be deduced from the storage modulus values alone. In the case of larger fillets, $G'$ covers the range 6000 to 10000~Pa for both undercooked and well-cooked samples. In the case of smaller "tail" fillets, $G'$ ranges from 2000 to 6000~Pa for undercooked, well-cooked and overcooked fish. To understand the absence of correlation between the cooking aspect of the fillets and the measured storage modulus, we need to keep in mind the complex structure of the fish muscles. One of the criterion for the visual evaluation of the cooking is the breakage of the fillets into flakes. In raw fish, the muscle fibers are linked to the myocommata and are hard to detach (Fig. \ref{figure_samples}(c)). When cod is cooked, the myocommata tends to break and myotome flakes are obtained \cite{1982_Kent}, as can be seen in Fig. \ref{figure_samples}(d). During the preparation of the samples for the rheometry measurement, each 16mm-diameter disk-shaped sample is cut in one flake (see Fig. \ref{figure_samples}(e)).
In the undercooked fillets where the flakes have not formed, we have avoided to include any white membrane in the final sample. Therefore, the breakage of the myocommata, though it affects the appreciation of the cooking, does not impact the rheometry measurements.

The second main visual change of the fillets during cooking is their color. Unsalted samples become gelly-like and slightly translucent during cooking. Salted samples, on the other hand, take a white color, very similar to ordinary cooked cod. The white color appears first on the top of the fillets, where the salt has been spread, and seems to extend toward the center of the fish during cooking. In addition, brown surfaces sometimes locally appear during cooking (Fig. \ref{figure_samples}(d)). Surprisingly, this clear visual difference between salted and unsalted \textit{Lutefisk} is not reflected by the rheological measurements. The myotomes are composed of muscle fibers and connective tissue. Cooking food removes water, denatures the proteins (i.e. they loose irreversibly their shape) and break long molecules into smaller ones \cite{2000_Barham}. It has been observed in the case of meat that the denaturation of the proteins in the muscle fibers leads to a contraction, and thus to rigidification of the fibers, whereas denaturation of the collagen in the connective tissue softens it \cite{2000_Barham}. This process is observable in a long cooked beef stew, for instance, where well separated muscle fibers are obtained. Therefore, the final softness of the food results from the combination of the rigidities of the fibers and of the connective tissue. Our experimental results show that in the case of \textit{Lutefisk}, rigidification of the fibers and softening of the connective tissue result together into a constant rigidity of the myotomes.

\section{Conclusions}

\textit{Lutefisk} is a traditional Norwegian Christmas dish, which is challenging to cook at home. Overcooking leads to a non-appetizing soft gelly-like meat texture. The goal of this paper is to compare the consumer feeling with rheological measurements, namely the storage and loss modulii. Contrarily to our expectations, the modulii were not significantly affected by the cooking time and temperature, nor by the salt content. Although we observed visually a modification of the aspect of the fish during cooking, this change was not reflected in the rheological measurements. The storage modulus depended predominantly on the type of the fish fillets: average value for the large fillets was twice as much as the average value for the smaller "tail" fillets, showing the essential role of the size of the fish piece on the final properties of the dish. In addition, early-season \textit{Lutefisk} was found to have a smaller elastic modulus than in the main season.

In addition, we have noted that we did not manage to overcook the large fillets, by doubling the cooking time or by increasing the temperature. The smaller "tail" fillets, on the other hand, could easily overcook. This shows that the better the quality of the fillet, the easier it is to obtain a well-cooked \textit{Lutefisk}. Our results indicate that it is the thickness of the fish piece, i.e. the fillet size, which determines if the fish can be cooked to perfection. We suspect that much of the commercially available \textit{Lutefisk} is from fish of lower size, i.e. lower quality, hampering the possibility for hobby chefs to succeed the preparation. Our experiments allow us to suggest a methodology for cooking the smaller \textit{Lutefisk} fillets. Smaller "tail" fillets gave very satisfactory cooking at 60$^o$C, so decreasing cooking temperature and increasing cooking time seems a promising way to proceed and may help to mitigate food waste.

These results give some insight into some of the parameters that affect the elastic modulus of cooked \textit{Lutefisk}, but there are many other aspects of its phase transition that would be interesting to further explore in the future.

\begin{acknowledgments}
We wish to thank Morton Korsnes and SiO for supporting our work by giving us access to the kitchen where we prepared the fish, and Stéphane Poulain for fruitful discussions. We acknowledge support from the Research Council of Norway and the National Science Foundation of USA under project ‘‘Multi-scale, Multi-phase Phenomena in Complex Fluids for the Energy Industries’’, Award Number 1743794.  
\end{acknowledgments}

\section*{Data Availability Statement}

The data that support the findings of this study are available from the corresponding author upon reasonable request.

\bibliography{Bibliography_Lutefisk}

\end{document}